\documentclass[twocolumn,showpacs,twoside,prl,superscriptaddress]{revtex4}
\usepackage{amsmath,amssymb,amsfonts,amsthm,latexsym,dsfont,mathrsfs,doi,hyperref,graphicx,braket,color,xcolor,bm,dcolumn,upgreek}

\begin{document}

\title{
	Connection between the N00N State and a superposition of SU(2) Coherent States
	}

\author{Barry C.\ Sanders}
\affiliation{%
	Institute for Quantum Science and Technology, University of Calgary, Alberta T2N~1N4, Canada%
	}
\affiliation{%
	Hefei National Laboratory for Physical Sciences at the Microscale
	and Department of Modern Physics.
	University of Science and Technology of China, Anhui 230026, China%
	}
\affiliation{%
	Shanghai Branch,
	CAS Center for Excellence and Synergetic Innovation Center
		in Quantum Information and Quantum Physics,
	University of Science and Technology of China, Shanghai 201315, China%
	}
\affiliation{%
	Program in Quantum Information Science, 
	Canadian Institute for Advanced Research,\\ Toronto, Ontario M5G~1Z8, Canada%
	}
\author{Christopher C. Gerry}
\affiliation{%
	Department of Physics, 
	Lehman College of the City University of New York, Bronx, New York 10468, USA%
	}

\date{\today}
\begin{abstract}
The N00N state,
which was introduced as a resource for quantum-enhanced metrology,
is in fact a special case of a superposition of two SU(2) coherent states.
Here we show explicitly the derivation of the N00N state from the superposition state.
This derivation makes clear the connection between these seemingly disparate states
and shows how the N00N state can be generalized to a superposition of SU(2) coherent states.
\end{abstract}

\maketitle

The N00N state~\cite{BKA+00,Dow08,GLM11}
\begin{equation}
\label{eq:N00N}
	|N00N\rangle
		:=\frac{1}{\sqrt{2}}
			\left(\text{e}^{-\text{i}\phi}|N\rangle\otimes\ket{0}
				+\text{e}^{\text{i}\phi}\ket{0}\otimes|N\rangle\right),
\end{equation}
is significant for its quantum-enhanced metrological application.
The N00N state~(\ref{eq:N00N}) is often defined for the case~$\phi=0$,
but generalizing to include~$\phi\neq 0$~\cite{DD07}
(see also \cite{WS03,WLD07})
retains the state's benefits and enables more general
ways to produce the N00N state than in the restrictive case with $\phi=0$.
In Eq.~(\ref{eq:N00N}) we have denoted
\begin{equation}
	|N\rangle:=\frac{\left(\hat{a}^\dagger\right)^N}{\sqrt{N!}}\ket{0}
\end{equation}
as the $N$-photon Fock state,
and the $N=0$ Fock state corresponds to the vacuum state.

The N00N state's application is to estimating a two-channel interferometric phase shift~$\phi$
with a phase precision that scales as
\begin{equation}
\label{eq:qscaling}
	\Delta\phi\propto 1/N;
\end{equation}
that is, the uncertainty of the phase estimate is inversely proportional to the number of photons
injected into a two-channel interferometer (v.g., Sagnac, Michelson or Mach-Zehnder configurations),
assuming a single use of each photon.
This scaling~(\ref{eq:qscaling}) reaches the quantum limit and quadratically improves on the
standard quantum limit, or shot-noise limit,
$\Delta\phi\propto 1/\sqrt{N}$.
 
Dowling's comprehensive review of the N00N state says that the N00N state was
first introduced in 1989~\cite{Dow08}.
He writes,
	``The N00N state was first discussed in 1989 by Barry Sanders,
	who was particularly interested in the Schr\"{o}dinger-cat aspect
	and how that affected quantum decoherence.''
This sentence is not quite correct as stated,
but the N00N state and the SU(2) ``Schr\"{o}dinger cat'' state~\cite{San89} do have a connection,
which is what we discuss in the balance of this brief report.


The 1989 paper presents an analysis of nonlinear spin dynamics and shows, 
at a particular point in the evolution
and for an initial spin
[or SU(2)] coherent state~\cite{Rad71,ACGT72,Per72},
the emergence of a superposition of two spin coherent states differing in phase by~$\pi$.
We must first choose state parameters that place the spin coherent states of the superposition antipodally on the equator of the corresponding Bloch sphere.
However, the superposition obtained is not a N00N state
in the usual Schwinger realization of the angular momentum algebra and states in terms of two sets of Bose operators~\cite{CMM+06}.
Rather, we require an applied rotation by~$\pi/2$
along an axis in the equatorial plane of the Bloch sphere in order to produce the required superposition of north-south angular momentum states.
It is this superposition, when it is converted to the Bose basis, that will be the N00N state. 

To understand the superposition of two SU(2) coherent states,
we first review the nature of the coherent state followed by its boson realization.
The SU(2) group is generated by the $\mathfrak{su}(2)$ algebra
spanned by the ladder operators~$\hat{J}_\pm$ (raising and lowering respectively)
and the weight (or Cartan) operator~$\hat{J}_z$.
The algebra is given by the commutator relations
\begin{equation}
\label{eq:algebra}
	\left[\hat{J}_\pm,\hat{J}_z\right]=\mp\hbar\hat{J}_\pm,\,
	\left[\hat{J}_+,\hat{J}_-\right]=2\hbar\hat{J}_z.
\end{equation}
The Casimir operator is the quadratic operator~$\hat{J}^2$,
which commutes with each operator $\hat{J}_\pm$ and~$\hat{J}_z$.

From~$\hat{J}_\pm$,
we can define the self-adjoint operators
\begin{equation}
\label{eq:JxJy}
	\hat{J}_x:=\frac{\hat{J}_++\hat{J}_-}{2},\;
	\hat{J}_y:=\frac{\hat{J}_+-\hat{J}_-}{2\text{i}}.
\end{equation}
Then
\begin{equation}
\label{eq:xycommutator}
	\left[\hat{J}_x,\hat{J}_y\right]
		=\frac{\text{i}}{2}\left[\hat{J}_+,\hat{J}_-\right]
		=\hbar\hat{J}_z,
\end{equation}
and the commutator holds for cyclic permutations of the indices $x$, $y$, and~$z$.

The Casimir operator spectrum is
\begin{equation}
\label{eq:spectrum}
	\{j=0,1/2,1,3/2,\dots\},
\end{equation}
and the weight spectrum is
\begin{equation}
	\{-j\leq m\leq j\}
\end{equation}
in integer steps.
The basis is $\{\ket{j,m}\}$ such that
\begin{equation}
\label{eq:J2j,m}
	\hat{J}^2\ket{j,m}_z=j(j+1)\ket{j,m}_z,\;
	\hat{J}_z\ket{j,m}_z=m\ket{j,m}_z.
\end{equation}
The parameter~$j$ designates an irreducible representation (irrep)
for which~$\hat{J}^2=j(j+1)\mathds{1}$ over a $(2j+1)$-dimensional Hilbert space.

The SU(2) coherent state is given by~\cite{ACGT72,Per72}
\begin{align}
\label{eq:SU2coherentstate}
	\ket{j,\gamma}
		=&R(\gamma)\ket{j,-j}_z
				\nonumber\\
		=&\left(1+|\gamma|^2\right)^{-j}
			\sum_{m=-j}^{j}\begin{pmatrix}2j\\j+m\end{pmatrix}^{1/2}
			\gamma^{j+m}\ket{j,m}_z
\end{align}
where the ``rotation operator'' $R(\gamma)$ is given as
\begin{equation}
\label{eq:rotationoperator}
	R(\gamma)
		=\exp\left\{-\frac{1}{2}\theta\left(
			\hat{J}_+\text{e}^{-\text{i}\varphi}
			-\hat{J}_-\text{e}^{\text{i}\varphi}\right)\right\},
\end{equation}
and where
\begin{equation}
\label{eq:stereographic}
	\gamma=\text{e}^{\text{i}\varphi}\tan(\theta/2),\;
	0\leq\varphi\leq 2\pi,\;
	0\leq\theta\leq\pi.
\end{equation}
The parameters~$\varphi$ and~$\theta$ 
are respectively the azimuthal and polar coordinates on the Bloch sphere,
and~$\gamma$ is the complex coordinate on the plane obtained by a stereographic projection of the sphere used to represent the rotation operator~\cite{San89}.
Each SU(2) coherent state~(\ref{eq:SU2coherentstate})
can be identified with a point on a sphere 
(the values of~$\theta$ and~$\varphi$) or,
equivalently, with a point on the plane obtained with a stereographic project of the sphere.

Let us now consider some extreme cases of the SU(2) coherent state
of Eq.~(\ref{eq:SU2coherentstate}).
For $\theta=0$ and with an arbitrary value of~$\phi$,
we have~$\gamma=0$, so $\ket{j,\gamma=0}=\ket{j,j}_z$.
The other extreme is for $\theta\to\pi$,
again for~$\phi$ arbitrary,
in which case $\gamma\to\infty$
such that $\ket{j,\gamma\to\infty}=\ket{j,-j}_z$.
These states are extreme in the sense that they are antipodal on the Bloch sphere.

As another case,
we consider $\ket{j,\gamma=\pm\text{i}}$,
which corresponds to $\theta=\pi/2$,
the equator of the Bloch sphere, and $\phi=\pm\pi/2$.
These extreme cases are those of the angular momentum state basis
where the operator~$\hat{J}_y$ is diagonal;
i.e., $\ket{j,\pm\text{i}}=\ket{j,\pm j}_y$.
The two sets of extreme states are related by the $\pi/2$ rotation
about the $x$-axis
\begin{equation}
	R_x(\pi/2)\ket{j,\pm j}_y=\ket{j,\pm j}_z,
\end{equation}
where
\begin{equation}
	R_x(\pi/2)=\exp\left[-\text{i}(\pi/2)\hat{J}_x\right].
\end{equation}
We use these results presently.

We first need to generate a superposition of the SU(2) coherent states.
As was shown in Ref.~\cite{San89},
the nonlinear Hamiltonian
\begin{equation}
\label{eq:Hj}
	\hat{H}_j
		=\omega\hat{J}_z+\frac{\lambda}{2j}\hat{J}_z^2
\end{equation}
acting for one quarter of the period 
$\tau_j=4\pi j/\lambda$ on the state $\ket{j,\gamma}$
results in the superposition state
\begin{align}
\label{eq:nonlinearevolution}
	\exp&\left(-\text{i}\hat{H}_j\tau_j/4\right)\ket{j,\gamma}
				\nonumber\\
		=&\frac{1}{\sqrt{2}}\left[\text{e}^{-\text{i}\pi/4}\ket{j,\gamma}
			+(-1)^j\text{e}^{\text{i}\pi/4}\ket{j,-\gamma}\right],
\end{align}
a spin-coherent-state form of the Schr\"{o}dinger cat state.

If we now set $\gamma=\text{i}$,
we can use the above results to obtain
\begin{align}
\label{eq:nonlinearevolutioni}
	\exp&\left(-\text{i}\hat{H}_j\tau_j/4\right)\ket{j,\text{i}}
				\nonumber\\
		=&\frac{1}{\sqrt{2}}
			\left[\text{e}^{-\text{i}\pi/4}\ket{j,\text{i}}
			+(-1)^j\text{e}^{\text{i}\pi/4}\ket{j,-\text{i}}\right]
				\nonumber\\
		=&\frac{1}{\sqrt{2}}
					\left[\text{e}^{-\text{i}\pi/4}\ket{j,j}_y
			+(-1)^j\text{e}^{\text{i}\pi/4}\ket{j,-j}_y\right].
\end{align}

We now perform a~$\pi/2$ rotation about the $x$-axis to obtain
\begin{align}
\label{eq:nonlinearevolutionz}
	R_x\left(\frac{\pi}{2}\right)&\exp\left(-\text{i}\hat{H}_j\frac{\tau_j}{4}\right)\ket{j,j}_y
					\nonumber\\
		&=R_x\left(\frac{\pi}{2}\right)\frac{1}{\sqrt{2}}
			\left[\text{e}^{-\text{i}\pi/4}\ket{j,j}_y
			+(-1)^j\text{e}^{\text{i}\pi/4}\ket{j,-j}_y\right]
					\nonumber\\
		&=\frac{1}{\sqrt{2}}
			\left[\text{e}^{-\text{i}\pi/4}\ket{j,j}_z
			+(-1)^j\text{e}^{\text{i}\pi/4}\ket{j,-j}_z\right].
\end{align}
The right-hand side of the second line of this equation constitutes a N00N state
when expressed in the two-mode oscillator basis,
which is related to the angular momentum basis in which the
operator~$\hat{J}_x$ is diagonal,
as per the Schwinger realization of the $\mathfrak{su}$(2) algebra that will be discussed below.
This is the state we want. Note that it is not contained within the SU(2) superposition state,
rather that it is obtained from the state with extreme parameters by the rotation about the $x$-axis.

For completeness, we consider the left-hand side of Eq.~(\ref{eq:nonlinearevolutionz}),
which we write as
\begin{align}
		R_x\left(\frac{\pi}{2}\right)&\exp\left\{-\text{i}\hat{H}_j\frac{\tau_j}{4}\right\}\ket{j,j}_y
					\nonumber\\
			=&R_x\left(\frac{\pi}{2}\right)\exp\left\{-\text{i}\hat{H}_j\frac{\tau_j}{4}\right\}
				R_x\left(-\frac{\pi}{2}\right)R_x\left(\frac{\pi}{2}\right)\ket{j,j}_y
				\nonumber\\
		=&R_x\left(\frac{\pi}{2}\right)\exp\left\{-\text{i}\hat{H}_j\frac{\tau_j}{4}\right\}
				R_x\left(-\frac{\pi}{2}\right)\ket{j,j}_z.
\end{align}
Then we have
\begin{equation}
	R_x\left(\frac{\pi}{2}\right)\exp\left\{-\text{i}\hat{H}_j\frac{\tau_j}{4}\right\}R_x\left(-\frac{\pi}{2}\right)
		=\exp\left\{-\text{i}\hat{H}'_j\frac{\tau_j}{4}\right\}
\end{equation}
where
\begin{align}
\label{eq:H'j}
	\hat{H}'_j
		:=&R_x\left(\frac{\pi}{2}\right)\hat{H}_jR_x\left(-\frac{\pi}{2}\right)
		=\omega\hat{J}_y+\frac{\lambda}{2j}\hat{J}_y^2,
\end{align}
and where we have used the relations
\begin{equation}
\label{eq:Jrelations}
	R_x\left(\frac{\pi}{2}\right)\hat{J}_zR_x\left(-\frac{\pi}{2}\right)=\hat{J}_y,\;
	R_x\left(\frac{\pi}{2}\right)\hat{J}_z^2R_x\left(-\frac{\pi}{2}\right)=\hat{J}_y^2.
\end{equation}
Thus, we can rewrite Eq.~(\ref{eq:nonlinearevolutionz}) as
\begin{align}
\label{eq:righthandside}
	\exp&\left\{-\text{i}\hat{H}'_j\frac{\tau_j}{4}\right\}\ket{j,j}_z
					\nonumber\\
		=&\frac{1}{\sqrt{2}}\left[\text{e}^{-\text{i}\pi/4}\ket{j,j}_z
			+(-1)^j\text{e}^{\text{i}\pi/4}\ket{j,-j}_z\right],
\end{align}
which is an interesting result in its own right, a group theoretically derived result.

The correspondence between the spin states and the coupled harmonic oscillators,
and thus the N00N states,
is made from the Schwinger realization of the $\mathfrak{su}$(2) algebra
in terms of a set of Bose operators:
\begin{equation}
\label{eq:Schwingerrealization}
	\hat{J}_+=\hat{a}^\dagger\hat{b},\;\hat{J}_-=\hat{a}\hat{b}^\dagger,\;
	\hat{J}_z=\frac{1}{2}\left(\hat{a}^\dagger\hat{a}-\hat{b}^\dagger\hat{b}\right)
\end{equation}
and
\begin{equation}
\label{eq:J0}
	\hat{J}_0=\frac{1}{2}\left(\hat{a}^\dagger\hat{a}+\hat{b}^\dagger\hat{b}\right),
\end{equation}
where~$\hat{J}_0$ is the Casimir operator
such that the square of the total angular moment is given by
\begin{equation}
	\hat{\bm J}^2
		=\hat{J}_x^2+\hat{J}_y^2+\hat{J}_z^2=\hat{J}_0\left(\hat{J}_0+\mathds{1}\right),
\end{equation}
and where
\begin{align}
	\hat{J}_x
		=&\frac{1}{2}\left(\hat{J}_++\hat{J}_-\right)
		=\frac{1}{2}\left(\hat{a}^\dagger\hat{b}+\hat{a}\hat{b}^\dagger\right),
				\nonumber\\
	\hat{J}_y
		=&\frac{1}{2\text{i}}\left(\hat{J}_+-\hat{J}_-\right)
		=\frac{1}{2\text{i}}\left(\hat{a}^\dagger\hat{b}-\hat{a}\hat{b}^\dagger\right).
\end{align}
The angular momentum states $\ket{j,m}_z$
satisfy the eigenvalue relations
\begin{equation}
\label{eq:eigenvaluerelations}
	\hat{J}_0\ket{j,m}_z=j\ket{j,m}_z,\;
	\hat{J}_z\ket{j,m}_z=m\ket{j,m}_z,\;
\end{equation}
and thus correspond to the $a$- and $b$-mode Fock basis according to 
\begin{equation}
\label{eq:Fock}
	\ket{j,m}_z=\ket{j+m}_a\otimes\ket{j-m}_b.
\end{equation}
Two special cases are
\begin{equation}
\label{eq:2j002j}
	\ket{j,j}_z=\ket{2j}_a\otimes\ket{0}_b,\;
	\ket{j,-j}_z=\ket{0}_a\otimes\ket{2j}_b.
\end{equation}
Thus, the right-hand side of Eq.~(\ref{eq:nonlinearevolutionz}) is the N00N state
\begin{align}
\label{eq:N00Nmaker}
	\ket{\text{N00N}}
		=&\frac{1}{\sqrt{2}}
			\Big[\text{e}^{-\text{i}\pi/4}\ket{2j}_a\otimes\ket{0}_b
				\nonumber\\
			&+(-1)^j\text{e}^{\text{i}\pi/4}\ket{0}_a\otimes\ket{2j}_b\Big],
\end{align}
in agreement with the right-hand side of Eq.~(\ref{eq:N00N}) for $N=2j$,
$\phi=\pi/4$ and, for even~$N$ such that $(-1)^j=1$.
However, this result is a N00N state for odd~$N$ as well as even.

In summary, from superpositions of the $\mathfrak{su}$(2) coherent states~$\ket{j,\gamma}$
and~~$\ket{j,-\gamma}$
with extreme parameters that place them at antipodal positions on the Bloch sphere,
one can obtain the equivalent of N00N states \emph{provided}
that a $\pi/2$ rotation is implemented to align the superposition states along the $z$-axis as a final step.
We could just have easily have chosen $\gamma=1$ as the extreme parameter,
in which case the final step would be a $\pi/2$ rotation about the $y$-axis.

\acknowledgements
This research was supported by AITF, NSERC, CIFAR, and the China Thousand Talent Program.
We acknowledge valuable comments from Vahid Karimipour.

\bibliography{N00N}

\end{document}